\documentclass {aastex63}

\graphicspath{{./}{figures/}}

\received{January 1, 2018}
\revised{January 7, 2018}
\accepted{\today}
\submitjournal{ApJ}

\shorttitle{Approximate MHD Slow Body Mode Solutions in Photospheric Waveguides}
\shortauthors{A. A. Aldhafeeri et al.}

\begin{document}

\title{Comparison of exact and approximate MHD slow body mode solutions in photospheric waveguides}

\correspondingauthor{Anwar A. Aldhafeeri}
\email{aaaldhafeeri@kfu.edu.sa}

\author[0000-0003-2220-5042]{Anwar A. Aldhafeeri}
\affiliation{Mathematics and Statistic Department, Faculty of Science, King Faisal University, Al-Hassa, P.O. Box 400, Hofuf 31982, Saudi Arabia}

\author{Gary Verth}
\affiliation{Plasma Dynamics Group, School of Mathematics and Statistics, University of Sheffield, Hicks Building, Hounsfield Road, Sheffield, S3 7RH, UK}

\author{Viktor Fedun}
\affiliation{Plasma Dynamics Group, Department of Automatic Control and Systems Engineering, The University of Sheffield, Mappin Street, Sheffield, S1 3JD, UK}

\author{Matthew Lennard}
\affiliation{Plasma Dynamics Group, Department of Automatic Control and Systems Engineering, The University of Sheffield, Mappin Street, Sheffield, S1 3JD, UK}

\author[0000-0002-3066-7653]{I. Ballai}
\affiliation{Plasma Dynamics Group, School of Mathematics and Statistics, University of Sheffield, Hicks Building, Hounsfield Road, Sheffield, S3 7RH, UK}

\begin{abstract}
In this study we explore the possibility of simplifying the modeling of  magnetohydrodynamic (MHD) slow body modes observed in photospheric magnetic structure such as the umbrae of sunspots and pores. The simplifying approach assumes that the variation of the eigenvalues of slow body waves can be derived by imposing that the  longitudinal component of velocity with respect to the tube axis is zero at the boundary of the magnetic flux tube, which is in a good agreement with observations. To justify our approach we compare the results of our simplified model for slow body modes in cylindrical flux tubes with the model prediction obtained by imposing the continuity of the radial component of the velocity and total pressure at the boundary of the flux tube. Our results show that, to a  high accuracy (less than 1\% for the considered model), the conditions of continuity of the component of  transversal velocity and pressure at the boundary can be neglected when modelling slow body modes under photospheric conditions. 
\end{abstract}


\section{Introduction}

The correct identification  as well as the interpretation of magnetohydrodynamic (MHD) wave modes in a variety of magnetic structures is possibly one of the most important, yet most challenging, problems of solar physics. The study of waves in solar magnetic structures is important for many reasons. Waves can transport kinetic energy to the upper layers of the solar atmosphere, where it can be dissipated in the presence of steep gradients  in plasma or field parameters and converted to heat, therefore, contributing to the plasma heating \citep[see e.g.][]{Erdelyi2007,  Usmanov2016, oran2017, Tiwari2019, Cranmer2020}. 

On the other hand, MHD waves observed with high resolution can be used as a diagnostic tool for the properties of the plasma and magnetic field using a seismological approach. This technique involves the combination of theoretical models (dispersion relation, evolutionary equation, etc.) with high resolution observations that provide information on the damping time/length of waves, their periods, wavelength, amplitude, etc. to derive values for physical parameters that  cannot be determined by direct measurements (sub-resolution structure of the magnetic field, heating/cooling functions, transport coefficients, scale-heights, ionisation degree, optical depths, etc.). 

The derivation of an analytical dispersion relation depends on the geometry of the waveguide and coordinate system applied, assuming that the boundary of the waveguide varies smoothly with the radius and shows symmetry with the longitudinal axis. We should mention that a dispersion relation can be derived even in the case of a discontinuity separating two regions of different properties (see, e.g. \citet{Roberts1981a, Musielak2000, ballai2011A, Vickers2018}), however this setup is not entirely relevant to solar plasmas. In addition, in the absence of additional physical effects, waves will be dispersive only in waveguides that have a clear geometrical extent.

The magnetic field in the solar atmosphere tends to accumulate in geometrically well-defined structures of different strengths, from a few Gauss in the quiet Sun and corona to kilo Gauss fields (pores and sunspots). These magnetic structures show a temporal variability in  their shape, strength, stability, and density. 
Waves propagating along these magnetic structures have been studied assuming either a magnetic slab described in Cartesian geometry \citep{Roberts1981b,Marcu2005, Hornsey2014, Pascoe2016, Mather2018, Li2018, Skirvin2021}, or a magnetic flux tube \citep{Edwin1983,Fedun2006, verth2007, verth2010, Fedun2010, Jess2015} described in cylindrical geometry. These simple, yet instructive, models have the necessary symmetry that makes the derivation and analysis of a dispersion relation possible.

Probably the most studied model to describe MHD waves in solar magnetic structures is the magnetic cylinder that can provide quantitative and qualitative description to the myriad of waves that can appear in magnetic structures. According to the standard nomenclature \citep{Edwin1983}, waves can be categorised according to the way they perturb the symmetry axis of the tube. Sausage modes are axisymmetric waves that propagate such that the axis of the tube is not perturbed. In contrast, in the case of kink waves, the symmetry axis of the tube undergoes a swinging motion. Waves can also be categorised by the way they behave inside the cylinder in the transversal direction. Waves whose behaviour is oscillatory inside the waveguide are usually referred to as body waves. Waves that are evanescent inside the tube and have a maximum of their amplitude on the boundary of the tube as surface waves. Finally, waves can also be categorised according to their characteristic speeds. Accordingly, the possible magnetoacoustic waves can be slow or fast, and their phase speed is weakly (for slow waves) or strongly (for fast waves) influenced by dispersive effects. All the above categorisations are true as long as the magnetic field is along the interface separating the internal and external regions  (the discontinuity where the magnetic field is parallel to the discontinuity surface is often called a tangential discontinuity). When the background magnetic field has a tilt angle with respect to the interface, the interface becomes a contact discontinuity and the life-time of waves is shortened by lateral leakage of energy \citep[][]{ruderman2018,Vickers2018}.    

In reality, though, high-resolutions observations show that the cross-section of magnetic flux tubes is far from being circular (see e.g. \citet{Sobotka1999, Mathew2003, Ryutova2008, Solanki2003, Borrero2011, Schlichenmaier2016, Sobotka2017, Keys2018, Houston2018}). A very first attempt to consider non-circular waveguides (however, still maintaining a high degree of symmetry) is the study of waves in magnetic flux tubes of elliptical cross-section. The studies by \citet{Ruderman2003,Diaz2006, Morton2009, Morton2011, Guo2020} investigated the modifications of the waves' properties in such geometry. The results  of these investigations show that such structures can support the propagation of two kink waves polarised linearly along the minor and major axis of the flux tube, and in general waves with a non-zero azimuthal wavenumber are affected by the eccentricity of the elliptical shape.

Observationally waves can be categorised based on signatures such as symmetry or asymmetry. However, recent studies by \cite{Aldhafeeri2020} and \cite{Albidah2022} have shown that the morphology of waves (especially higher order modes) in sunspot umbrae are strongly affected by their cross-sectional shapes, and a wave diagnostic based on the perturbations of the intensity or Doppler velocity  might lead to wrong conclusions on the nature of waves if this is not taken into account. Hence, a more accurate depiction of theoretically modelled cross-sectional shapes is required to explain the eigenmodes of pores and sunspots. 

In particular, the effect of ellipticity of the magnetic flux tube on the patterns of MHD modes has been analysed in detail by \cite{Aldhafeeri2020}, who showed that for a high degree of ellipticity sausage modes corresponding to the $n = 0$ azimuthal wavenumber cannot be easily identified and  even solutions that correspond to the fluting mode with the azimuthal wavenumber $n = 3$ can be misinterpreted as a kink modes ($n = 1$) due to the similarities in their morphology. This result was the impetus for finding a new approach that clarifies and provides an accurate explanation of the type of observed wave mode patterns. It is worth noting here that most of the previous studies are based on the requirement of the continuity of the component of  radial (i.e. transversal) velocity and total pressure at the boundary of the waveguide. These conditions are a consequence of imposing that the normal and tangential components of the stress tensor are continuous. However, these conditions also constitute an obstacle in modelling the MHD modes with realistic cross-sectional shapes since mathematically, we do not have arbitrary coordinates to help us in the modeling process. Recent studies by \cite{Marco2022, Albidah2022} demonstrated the importance of taking into account the actual cross-sectional shape of sunspot umbrae in giving a correct interpretation of the observed modes, by taking the vertical  (i.e. longitudinal) velocity or density/pressure perturbations to be zero at the umbra-penumbra boundary, to be consistent with the observation data.

Motivated by the findings of \cite{Marco2022, Albidah2022} in this paper we demonstrate that the eigenvalues, and hence eigenfunctions, of slow body modes can be found without solving the full dispersion relation. The paper is structured as follows: in Section \ref{sec:2} we present the traditional way of determining the dispersion relation of MHD waves in circular waveguides, together with the governing equation of eigenvalues. Section \ref{sec3} is devoted to the comparative study of eigenvalues of slow body modes obtained using two different approaches. In Section \ref{sec:3} we compare the variation of eigenvalues near the boundary of the tube providing numerical and mathematical foundation for our approach. Finally, in Section \ref{sec:5}  we summarise and discuss our results.


\section{Governing equations}\label{sec:2}

The simplest, however illustrative, model to study wave propagation in solar magnetic flux tubes, e.g. coronal loops, sunspots, pores, prominence fibrils, etc., is the cylindrical magnetic flux tube model \citep{Wentzel1979, Wilson1979, Spruit1982, Edwin1983}. This model provides quantitative and qualitative description of possible magnetoacoustic waves that can propagate in such structures. The waves that a magnetic flux tube can support are used to explain the characteristics of  observed patterns of line of sight magnetic and velocity perturbations. One important consequence of waves confined to propagate in a geometrically well-defined structure is that waves become dispersive, i.e. their propagation speed depends on the wavelength of waves.

In the case of sunspots the magnetic field is predominantly vertical in the umbra, meaning that the cylindrical model of magnetic waveguide can be applied to this region. Waves studied with this model are mainly confined to the umbra region and are evanescent in the outer penumbra region. Under photospheric conditions it is customary to assume that the background plasma temperature, density and pressure in umbra is less than the in the penumbra and the  strength of the vertical magnetic field is greater in the umbra than the penumbra. These considerations lead to the particular ordering of the characteristic speeds $V_{Ai}$, $C_{Se} >C_{Si}$, $V_{Ae}$, where  $V_A$ denotes the Alfv\'en speed and  $C_S$ stands for the sound speed in the internal ({\it i}) and external ({\it e}) regions of the cylindrical flux tube. These particular conditions have important consequence for the types of MHD wave modes that can be supported by such a waveguide. Specifically, the MHD wave modes with the largest phase speeds, $V_{ph}$, propagating in the $z$ direction are surface modes with $V_{ph} \in (C_{Si},C_{Se})$, and these modes attain their maximum amplitude at the umbra/penumbra boundary. In addition, we found that such of a waveguide can also support slow body modes whose phase speeds satisfy $V_{ph}\in (C_{Ti},C_{Si})$, where $C_{Ti}$ is the tube speed inside the waveguide and will be defined later  (see Equation \ref{a19}). In structures with strong magnetic field, such as pores and sunspots, the values of the tube and sound speeds are very close to each other, and all slow body modes propagate in the same narrow band of the dispersion diagram (see \cite{Edwin1983}). Slow body modes attain their maximum amplitude inside the flux tube and they possess an oscillatory pattern in the transversal direction. Given the very narrow allowed propagation window of slow body modes we would like to show that it might be possible that solving the full dispersion relation will not be required, instead a much easier approach would give us a very accurate result. To show this, we postulate that slow modes show a negligible perturbation amplitude at the umbra/penumbra boundary and, therefore, it might be possible that the effect of the environment can be confidently neglected.


 We consider a straight magnetic cylinder oriented along the $z$-axis. In equilibrium state the plasma is characterised by its kinetic pressure $p_0$, and its density, $\rho_0$, while the equilibrium homogeneous magnetic field oriented along the symmetry axis of the magnetic cylinder is denoted by $B_0$. All equilibrium quantities are considered of be homogeneous and they are allowed to have a jump in their magnitude at the boundary of the waveguide situated at $r=a$. After linearising the ideal MHD equations the relation that describes the temporal and spatial evolution of small amplitude MHD perturbations reads \citep[see, e.g.][]{lighthill1960, Cowling1976, Roberts1981a, Aschwanden2005}: 
\begin{eqnarray}
\frac{\partial^4 \Delta}{\partial t^4}-(C_S^2+V_{A}^2)\frac{\partial^2}{\partial t^2}\nabla^2 \Delta+C_S^2 V_{A}^2\frac{\partial^2 }{\partial z^2}\nabla^2 \Delta =0. \label{a2} 
\end{eqnarray}
Here $\Delta=\nabla \cdot\mathbf{v}$ is the divergence of velocity perturbation, $V_{A}=B_0/\sqrt{ \mu_0 \rho_0}$ is the Alfv\'{e}n speed, $C_S=\sqrt{\gamma p_0/\rho_0}$ is the adiabatic sound speed, $\mu_0$ is the magnetic permeability of free space, $\gamma$ is the ratio of specific heats, and $\nabla^2$ is the Laplace operator. In a cylindrical coordinate system ($r$,\, $\theta$,\, $z$) the Laplace operator can be represented as
\begin{equation}
\nabla^2=   \frac{\partial^2 }{\partial r^2}+\frac{1}{r} \frac{\partial }{\partial r}+\frac{1}{r^2}\frac{\partial^2 }{\partial \theta^2}+\frac{\partial^2 }{\partial z^2}.
\end{equation}\label{eq:ch_1_1.4}
Since the plasma is homogeneous and infinitely extended in the azimuthal and longitudinal directions, we can Fourier analyse the problem and write the variable in Equation (\ref{a2}) as
\begin{equation}
\Delta= R(r)\exp[i(\omega t -n\theta-kz)], \label{Dc:ch_1_sol}    \end{equation}
where $n$ and $k$ are the wavenumber components in the azimuthal and longitudinal and  directions, $\omega$ is the angular frequency of oscillations (assumed here to be real quantity) and the function $R(r)$ is the amplitude of $\Delta$ that needs to be determined. 


After substituting the relation (\ref{Dc:ch_1_sol}) back in Equation (\ref{a2}), the governing equation becomes a Bessel differential equation (for details see \citet{Edwin1983}) whose solutions are well-known. After imposing the kinematic and dynamic boundary conditions (the continuity of the total pressure and the radial component of the velocity) at the radius of the tube, i.e. at $r=a$, the dispersion relations can be written as \citep[see, e.g.][]{Spruit1982,Edwin1983, Aschwanden2005}
\begin{equation}
\rho_i (k^2 V_{Ai}^2-\omega^2) m_e  \frac{ K^\prime_n (m_e a)}{K_n (m_e a)} = \rho_e (k^2 V_{Ae}^2-\omega^2)m_i  \frac{I_n^\prime(m_i a)}{I_n(m_i a)},
\label{eq:3}
\end{equation}
for surface waves $(m_i^2 > 0)$, and 
\begin{equation}
\rho_0 (k^2 V_{Ai}^2-\omega^2) m_e  \frac{ K^\prime_n (m_e a)}{K_n (m_e a)} = \rho_e (k^2 V_{Ae}^2-\omega^2)n_i  \frac{J_n^\prime(n_i a)}{J_n(n_i a)},
\label{eq:4}
\end{equation}
 for body waves $(m_i^2=-n_i^2 < 0)$. In the above relations the magnetoacoustic parameters, $m_i$ and $m_e$, are defined as
 \begin{equation}
 m_i^2 =\frac{(k^2C_{Si}^2-\omega^2)(k^2V_{Ai}^2-\omega^2)}{(C_{Si}^2+V_{Ai}^2)(k^2C_{Ti}^2-\omega^2)}.
 \label{eq:5}
  \end{equation}
\begin{equation}
  m_e^2 =\frac{(k^2C_{Se}^2-\omega^2)(k^2V_{Ae}^2-\omega^2)}{(C_{Se}^2+V_{Ae}^2)(k^2C_{Te}^2-\omega^2)}, \label{a17}
  \end{equation}
  and the  tube speeds $C_{Ti}$ and $C_{Te}$ given as
  \begin{equation}
 C_{Ti}^2= \frac{C_{Si}^2 V_{Ai}^2}{C_{Si}^2+V_{Ai}^2}, \quad C_{Te}^2 = \frac{C_{Se}^2 V_{Ae}^2}{C_{Se}^2+V_{Ae}^2}.\label{a19}
  \end{equation}
are the internal and external tube (cusp) speeds.  
Here all quantities with an index $i$ refer to region inside the tube ($r<a$) and the quantities with an index $e$ denote the external region ($r>a$). The functions $K_n$, $I_n$ are the modified Bessel functions, $J_n$ is the Bessel function of the first kind and the `prime' denotes the spatial derivative of these functions with respect to their argument. The dispersion relations provide the relation between the longitudinal wave number, $k$, (along the axis of the cylinder), azimuthal wave number, $n$, and the frequency, $\omega$. Given that these dispersion relations involve transcendental functions, simple analytical solutions can be obtained only for limiting cases, and for a complete solution one needs a numerical approach.

 The system of MHD equations also allows us to express the perturbations of various physical quantities in an explicit way. It is easy to show that the total internal pressure (i.e. the sum of kinetic and magnetic pressures), $P_i$, in the case of body waves is governed by 
\begin{equation}
    \frac{d^2 P_i}{d r^2}+\frac{1}{r}\frac{d P_i}{d r}+\left(n_i^2-\frac{n^2}{r^2}\right)P_i=0,
    \label{eq:6}
\end{equation}
and the relationship between the velocity perturbations and the total pressure can be written as
\[
v_{ri}=-\frac{i\omega}{\rho_{0i}(\omega^2-k^2V_{Ai}^2)}\frac{d P_i}{d r}, \quad v_{\theta i}=\frac{\omega n}{\rho_{0i}r(\omega^2-k^2V_{Ai}^2)}P_i,
\]
\begin{equation}
    v_{zi}=-\frac{ikC_{Si}^2}{\omega^2-k^2C_{Si}^2}\frac{1}{r}\frac{\partial }{\partial r}(rv_{ri})+\frac{kn^2C_{Si}^2\omega}{r^2\rho_{0i}(\omega^2-k^2V_{Ai}^2)(\omega^2-k^2C_{Si}^2)}P_i.
    \label{eq:7}
\end{equation}

Equation (\ref{eq:6}) is a typical Bessel differential equation whose solution that is bounded at $r=0$ can be written as
\[
P_i=A_iJ_n(n_ir),
\]
where $A_i$ is an arbitrary amplitude.
As a result, all eigenfunctions of the problem can be expressed in terms of Bessel functions, in particular, the $z$-component of the velocity can be written as \citep[see e.g.][]{Spruit1982},
\begin{equation}
v_{zi}=-A_i\frac{kC_{Si}^2\omega}{\rho_{0i}(\omega^2-k^2C_{Ti}^2)}J_n(n_ir).
\label{eq:8}
\end{equation}
According to the standard nomenclature waves that correspond to $n=0$ are the sausage modes, $n=1$ are the kink modes, while the modes corresponding to $n>1$ are referred to as fluting modes.

In what follows our investigation will focus on our statement according to which instead of deriving the dispersion relation and solving it, we will show that in the case of slow body modes it will suffice, with high accuracy, to solve the equation for $v_{zi}$ directly with imposing that on the boundary of the waveguide this quantity is zero (i.e. imposing Dirichlet boundary condition). Therefore, the eigenvalues will be solutions of the equation $J_n(n_ia)=0$, like in the case of waves on a circular membrane. 

\section{Slow body sausage and kink modes in photospheric flux tubes: variation of eigenvalues}
\label{sec3}

In the case a cylindrical waveguide the study by \cite{Edwin1983} showed that under photospheric conditions slow waves are restricted to propagate in a rather narrow window between the internal tube ($C_{Ti}$) and sound ($C_{Si}$) speeds and they show a monotonically increasing phase speed with increasing wavenumber (or decreasing wavelength). 

\begin{figure*}
\gridline{\fig{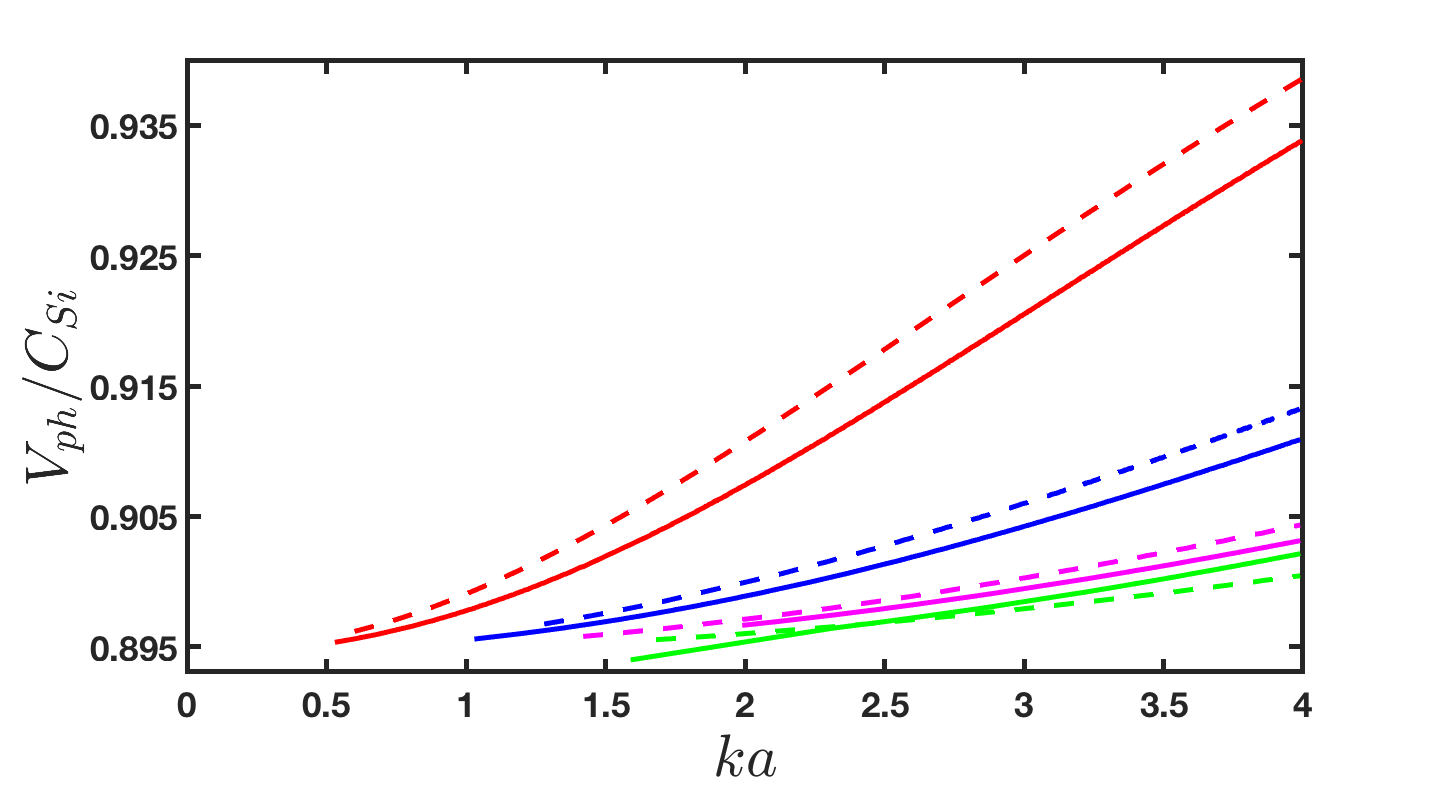}{0.5\textwidth}{(a)}
          \fig{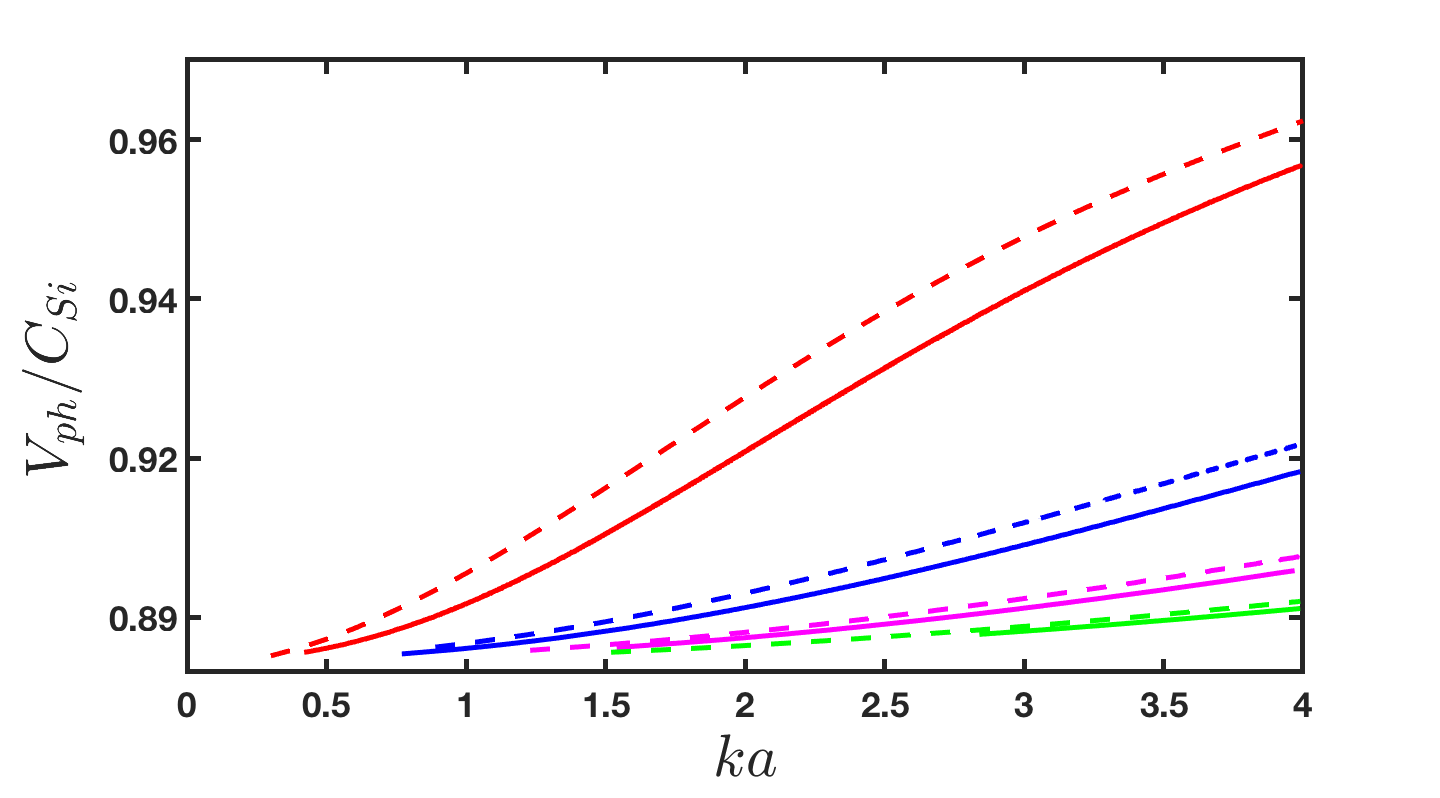}{0.5\textwidth}{(b)}
          }
\gridline{\fig{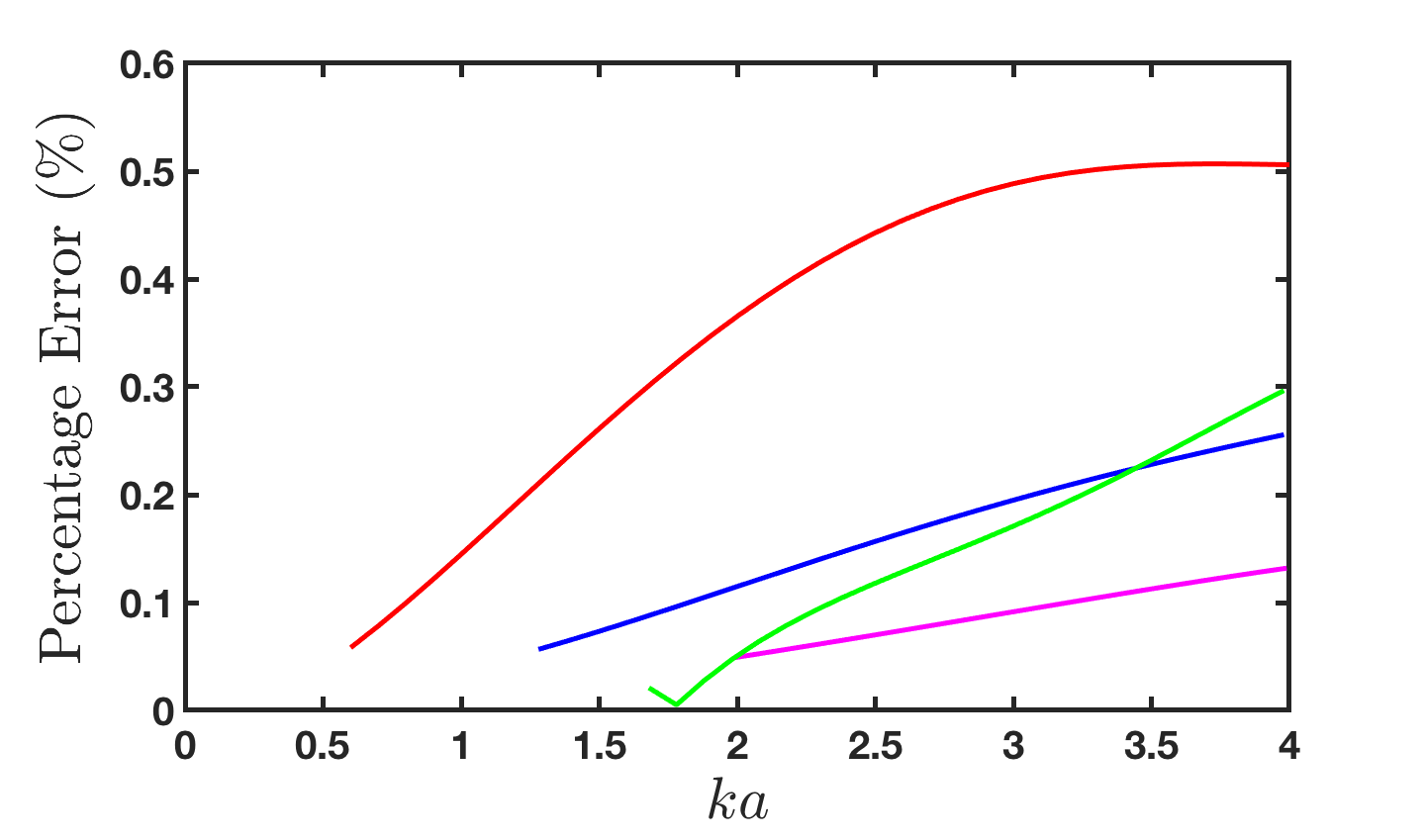}{0.5\textwidth}{(c)}
          \fig{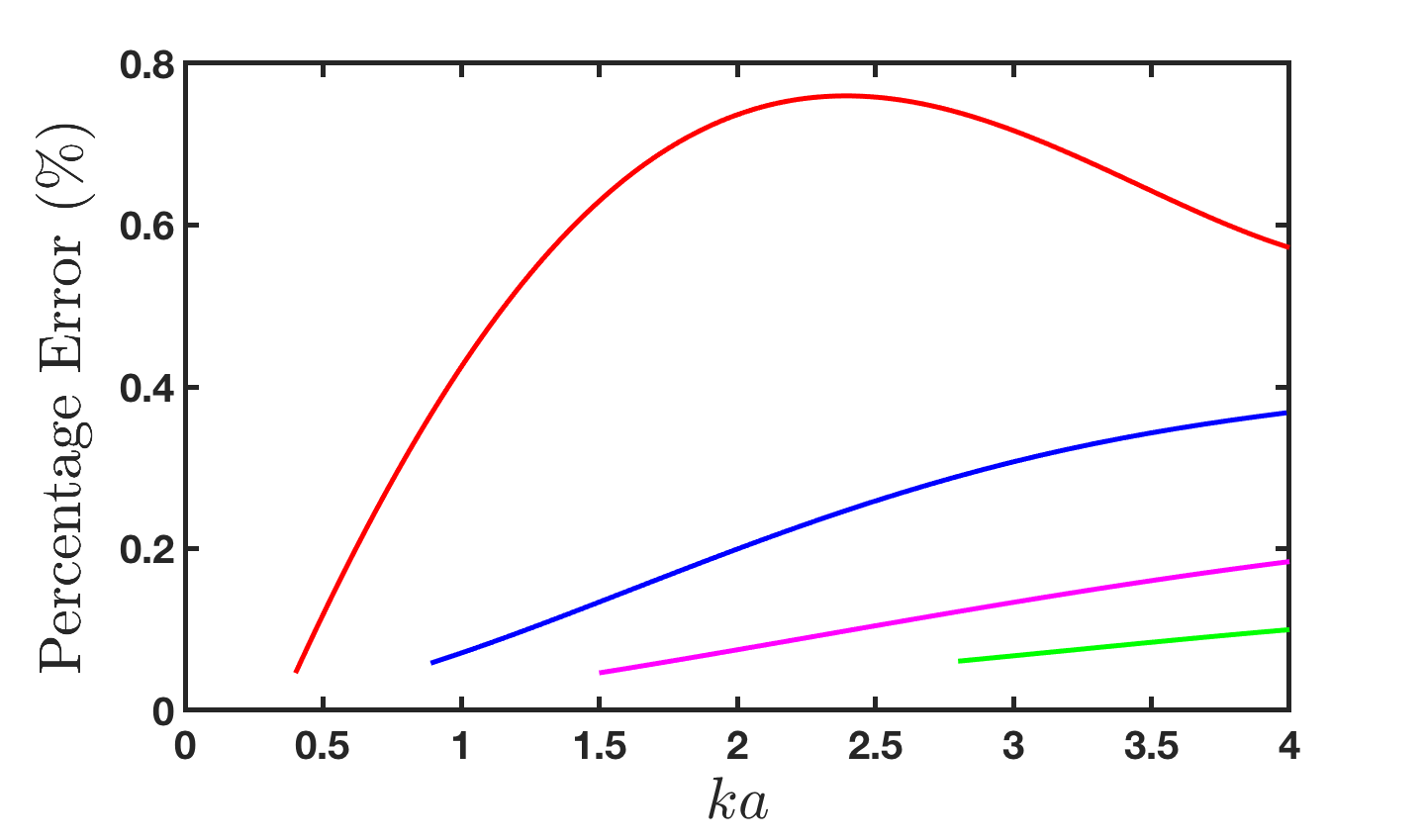}{0.5\textwidth}{(d)}
          }
\caption{The dispersion curves for slow body kink (panel a) and sausage (panel b) modes in a photospheric flux tube. The solid lines correspond to the variation of the phase speed of waves in units of the internal sound speed obtained based on the dispersion relation (Eq. \ref{eq:4}), while the dashed lines correspond to the solutions of the equation $v_{zi}=0$ on the boundary of the cylindrical waveguide, where the expression of $v_{zi}$ is given by Eq. (\ref{eq:8}). The different colours denote different wave harmonics, red stands for the fundamental modes, while blue, magenta and green lines denote subsequent higher harmonics. Panels (c) and (d) show the percentage error in finding the solutions of the dispersion relation Eq. (\ref{eq:4}) and the solutions obtained by assuming $v_{zi}=0$. The colour of the curves correspond to the wave harmonics of sausage and kink modes shown in panels (a-b).}
\label{fig:f1}
\end{figure*}

In the current Section we will compare the solutions we find for slow body sausage ($n=0$) and kink ($n=1$) modes by solving numerically the dispersion relation and by using the proposed simplified approach. For numerical calculations we adopt typical photospheric values for characteristic speeds, in particular we choose
$V_{Ae}= 0.5 C_{Si}$, $C_{Se} = 1.5C_{Si}$ and $V_{Ai} = 2C_{Si}$. The curves in panels (a) and (b) of Figure \ref{fig:f1} display the variation of the phase speed of slow body modes ($V_{ph}=\omega/k$) in units of the internal sound speed with respect to the dimensionless quantity $ka$ (here varying between 0 and 4) when the variation is obtained based on solving the full dispersion relation (\ref{eq:4}), solid lines and by solving $v_{zi}=0$ (dashed lines), where the expression of $v_{zi}$ is given by Eq. (\ref{eq:8}). It is clear from Figure \ref{fig:f1} that the two solutions are very close to each other not only in magnitudes, but also the trend. The fundamental sausage and kink modes have the highest propagating speed, higher harmonics propagate with lower phase speed. All modes are dispersive and the propagation speed increases with decreasing the wavelength. The variation of the dimensionless phase speed also shows that higher order modes are less and less affected by the change in the wavelength (or alternatively, by the size of the waveguide). The accuracy of the simplified method is shown by panels (c) and (d), where we plot the percentage error between the two approaches and the colours are chosen so that they match the modes shown in panels (a) and (b). For all modes the error is very small (less than 1\%) and error tends to diminish with increasing the order of the modes.  

The above results have important implications for the identification and modelling of waves in photospheric structures. First of all these results confirm that in the case of slow body modes it is a good approximation to consider that the longitudinal component of the velocity vanishes on the boundary. Since this velocity component is directly linked to the total pressure, this quantity also vanishes at the boundary. Obviously, this result does not mean that the whole velocity becomes zero, simple mathematical calculations evidence that here the radial component of the velocity has a maximum. 

Note that the assumption of the $v_{zi}= 0$ at the boundary is not valid if the value of $n_i a$ is less than the first zero of the Bessel function, $J_n(x)$, meaning that our approximate solution does not work accurately in a thin flux tube. In addition, our method also fails near the characteristic speeds, as at these values the argument of the Bessel function becomes zero or infinity, meaning that these speeds are degenerate values of the system. That is why the solutions we obtained in  Figure \ref{fig:f1} show a gap for long wavelengths. 

 To check how much our results vary with the range of Alfv\'en and sound speeds estimated from sunspot observations we use the characteristic values from \cite{Cho2017} who analysed 478 sunspots using the Helioseismic Magnetic Imager (HMI) onboard the Solar Dynamics Observatory (SDO). These authors found that the average relationship between the Alfv\'en and sound speeds in sunspot umbrae is $V_A=1.2C_S$. In order to investigate a wide spectrum of cases and to ensure the robustness of our results, we have repeated the analysis shown in Figure \ref{fig:f1} for $V_A=1.2C_S$ and $V_A=3C_S$. The results show that irrespective of the multiplier between the two characteristic speeds, the percentage error between the predictions of the full analysis and the simplified method remains less than 1\%. 

\section{Slow body sausage and kink modes in photospheric flux tubes: variation of eigenfunctions} \label{sec:3}

To further substantiate our approach for determining the dispersion curves for slow body modes in photospheric flux tubes, in this section we will discuss in detail the solutions obtained in cylindrical model following the standard approach of deriving dispersion relation with the help of matching the normal and tangential stresses at the boundary of the waveguide. The dispersion relation also allows us to express the eigenfunctions for different modes. Using the values of $\omega$ and $k$  based on the dispersion relations, we plot the variation of the $v_z$ velocity component under photospheric conditions in the neighbourhood of the waveguide's boundary.

As specified earlier, the set of linearised and ideal MHD equations in a cylindrical waveguide can be reduced so that the  $z$-components of the velocity inside and outside the waveguide are given by 
\begin{equation}
v_{zi}=- A_i\frac{k \omega C_{Si}^2}{\rho_{0i}V_{Ai}^2(\omega^2-k^2C_{Ti}^2)} J_n(n_ir), \nonumber
\end{equation}
\begin{equation}
v_{ze}=- A_e\frac{k \omega C_{Se}^2}{\rho_{0e}V_{Ae}^2(\omega^2-k^2C_{Te}^2)} K_n(m_er), \nonumber
\end{equation}
where $A_i$ and $A_e$ are arbitrary constants.

\begin{figure*}
\gridline{\fig{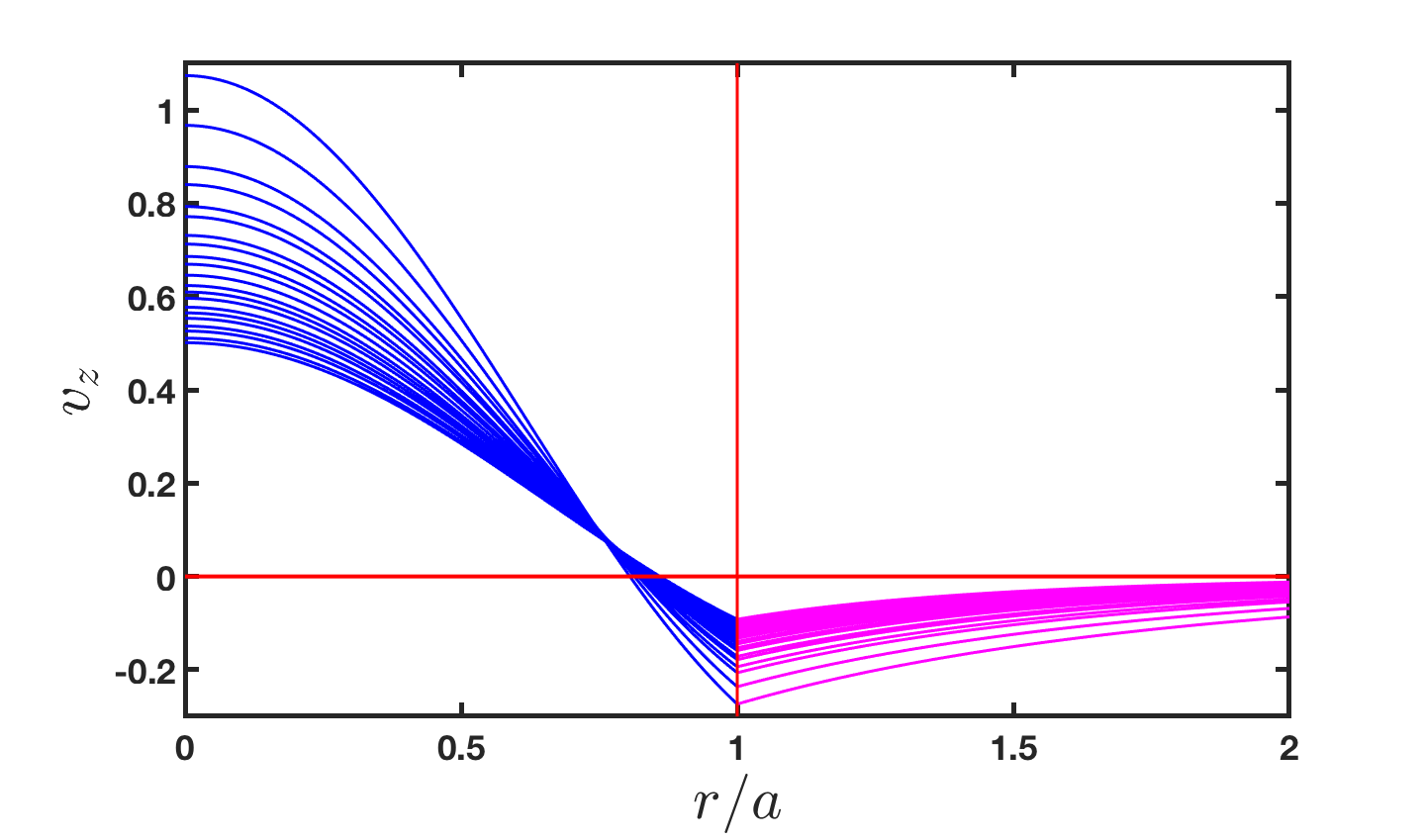}{0.35\textwidth}{(a)}
          \fig{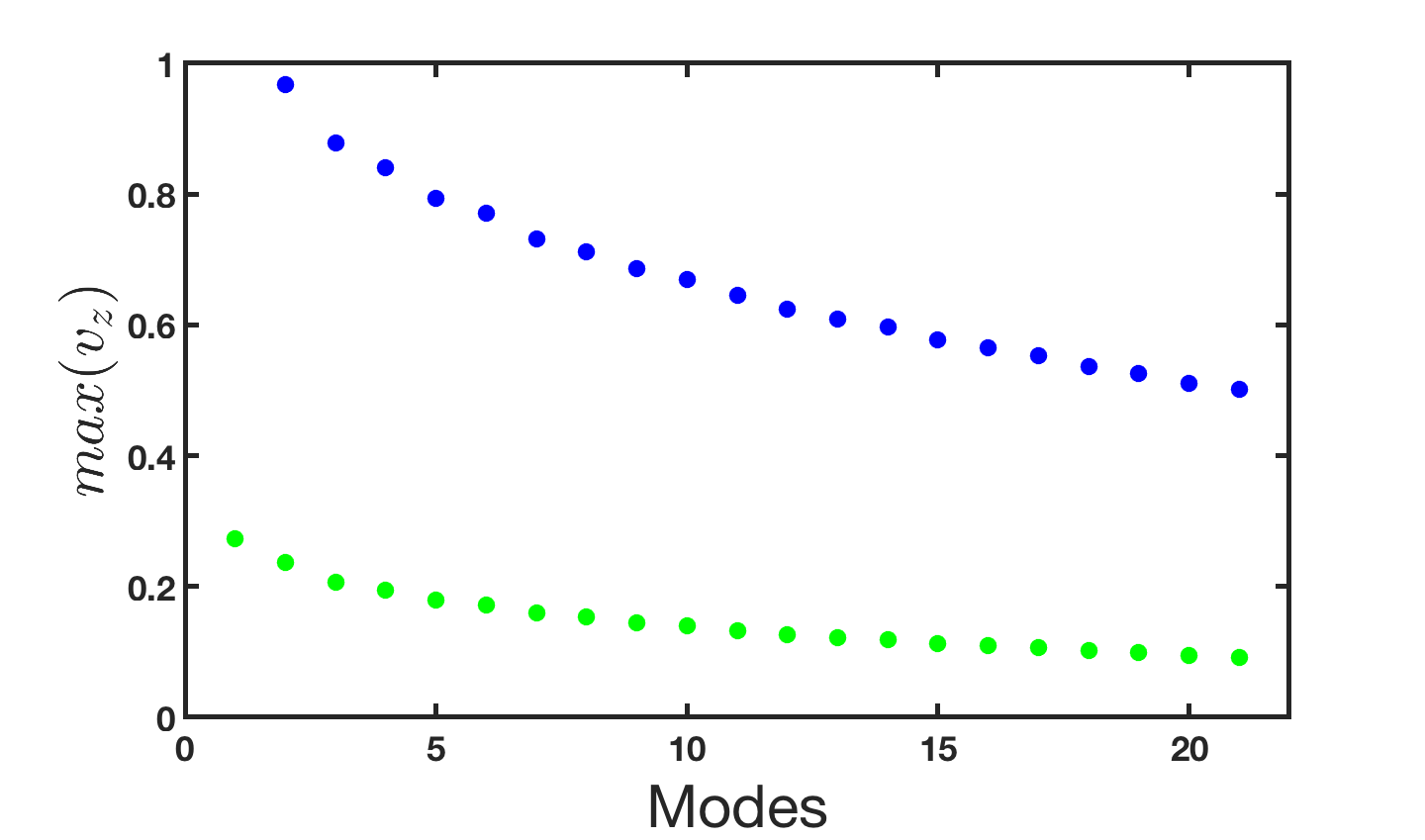}{0.35\textwidth}{(b)}
          \fig{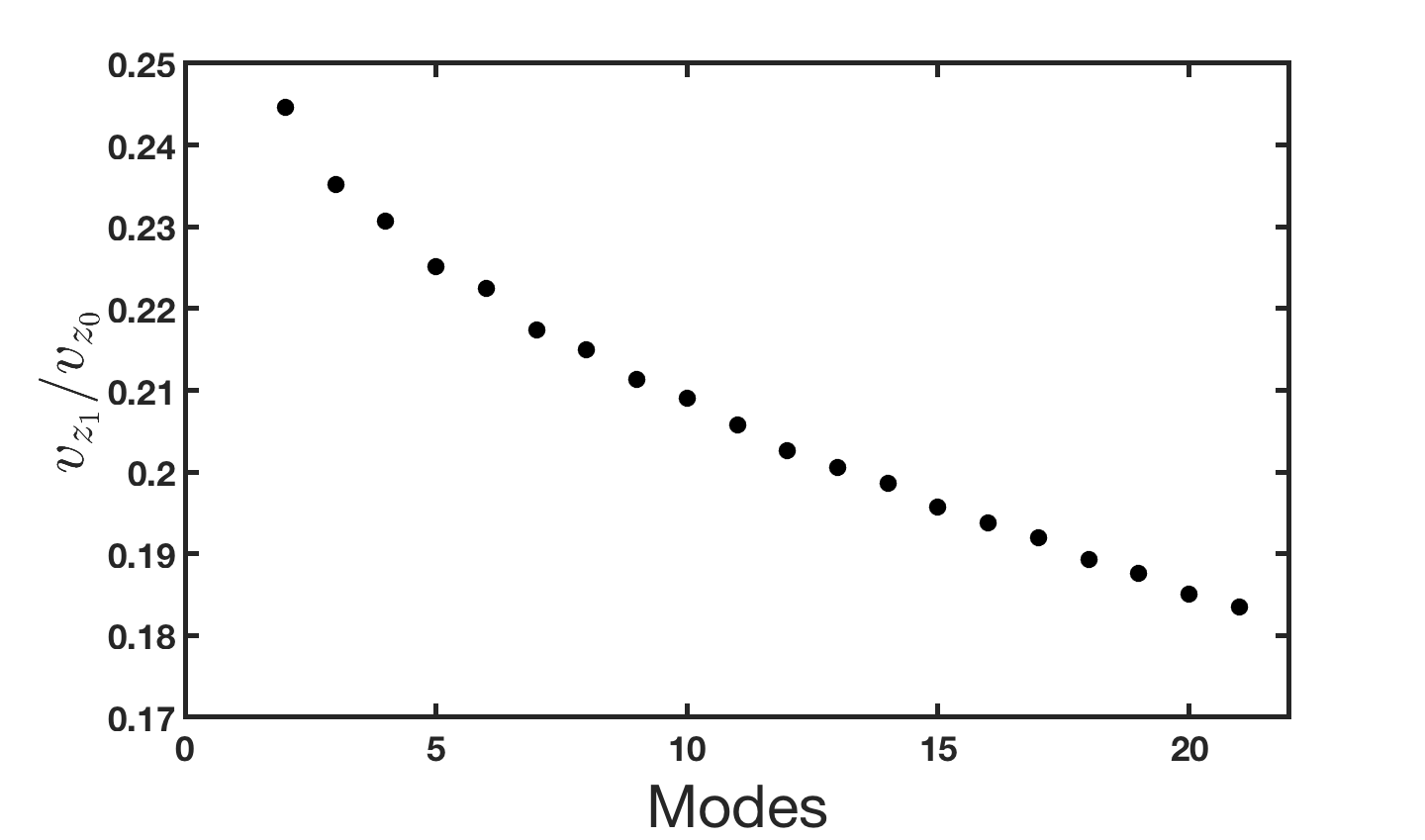}{0.35\textwidth}{(c)}
          }
\gridline{\fig{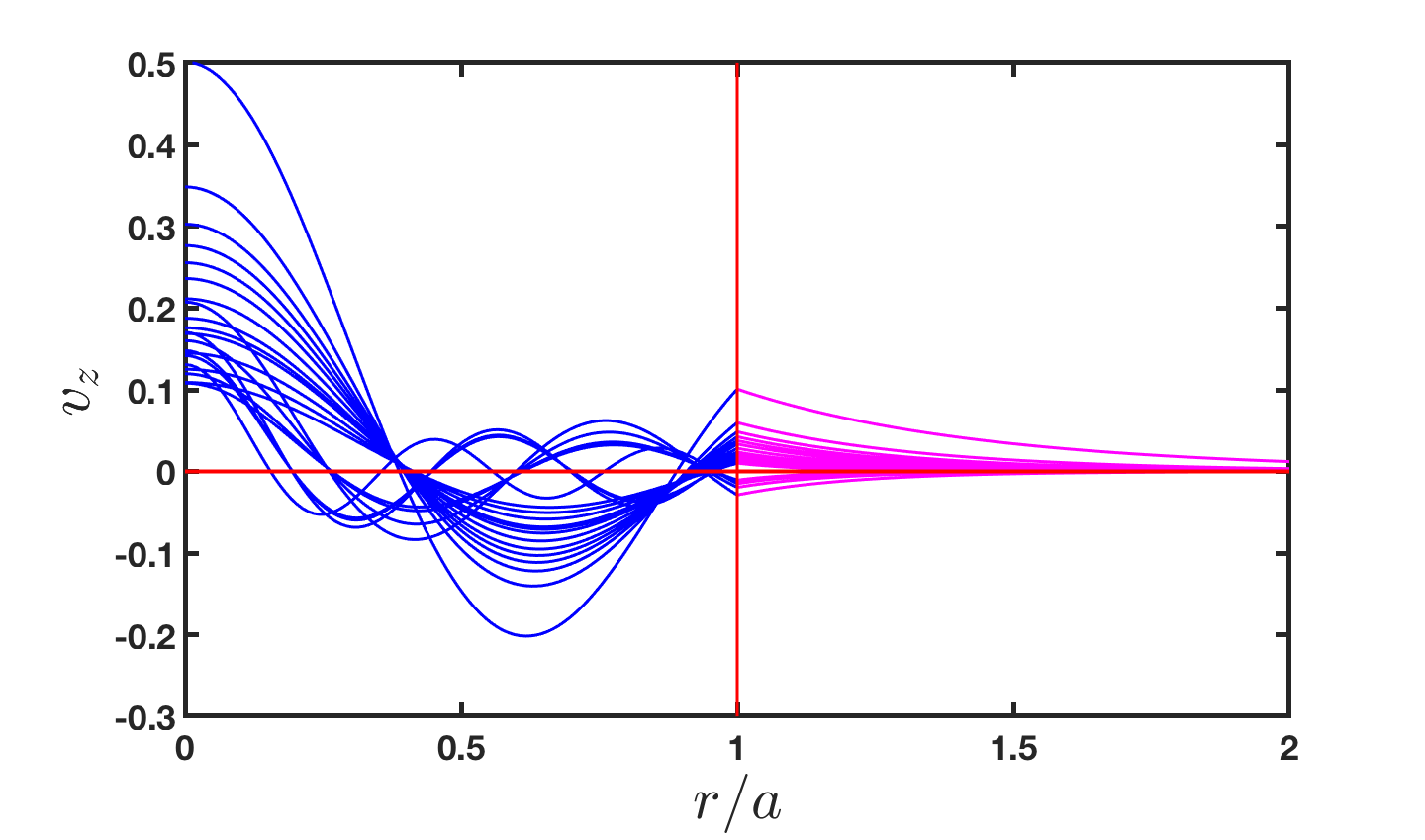}{0.35\textwidth}{(d)}
          \fig{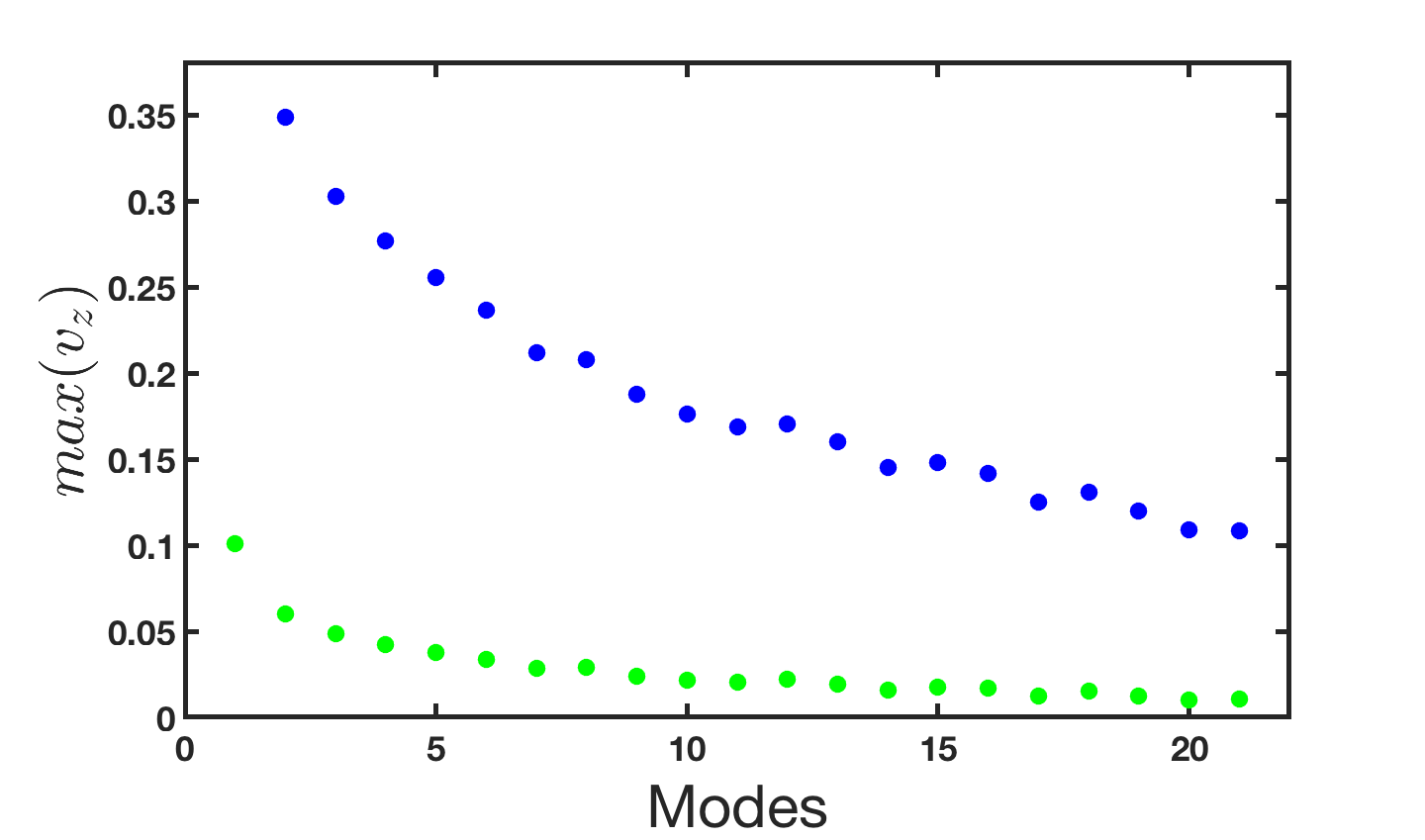}{0.35\textwidth}{(e)}
          \fig{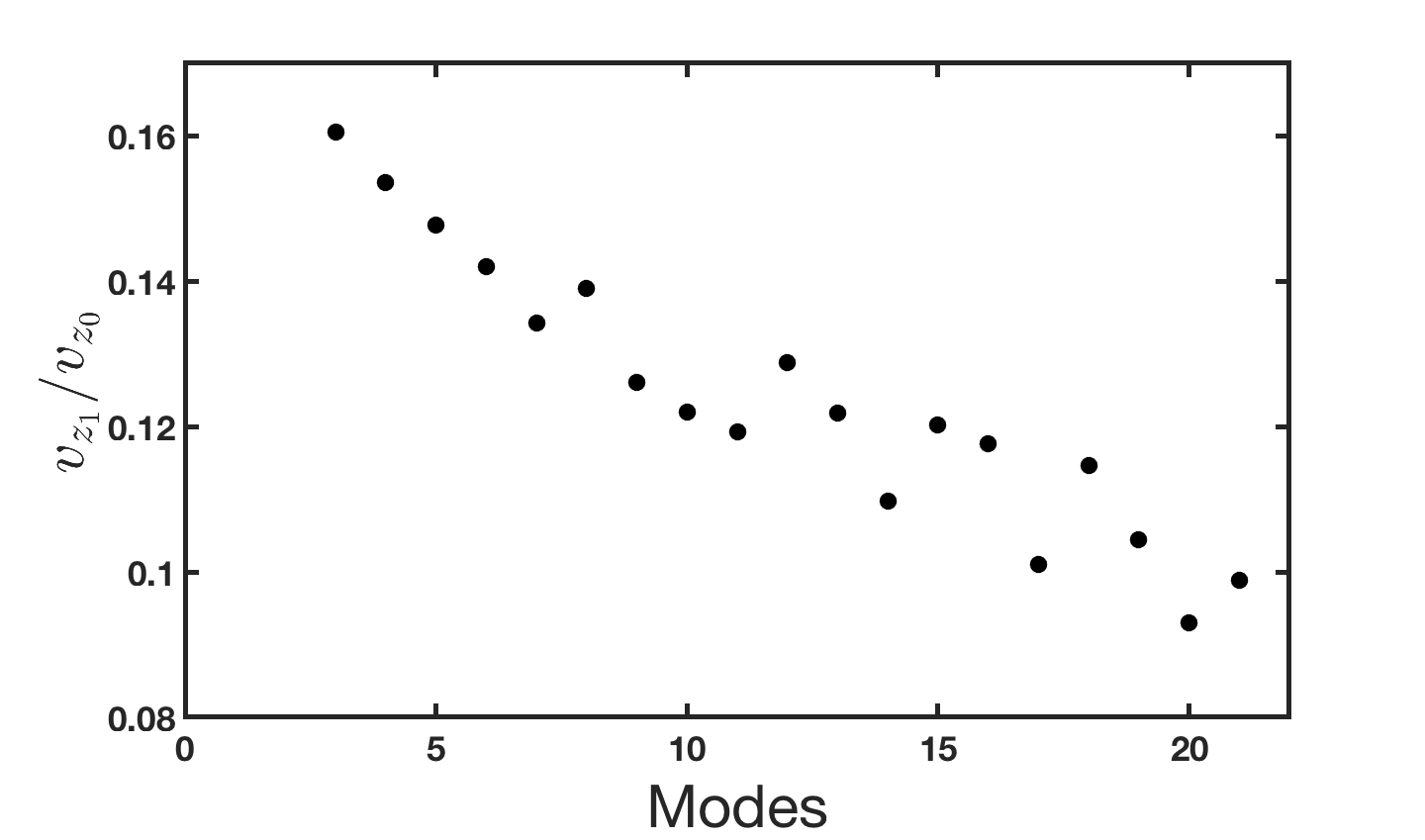}{0.35\textwidth}{(f)}
          }
\caption{The variation of the $z$-component of the velocity across the boundary of a cylindrical waveguide for slow sausage body modes under photospheric conditions. This eigenfuction is plotted for $an_i <4$ (panels a, b, and c) and for $ an_i \geq 4$ (panels d, e, and f), respectively. Panels (a) and (d) show the dependence of the amplitude of $v_z$ on the radial distance in units of the waveguide's radius inside (indicated in blue) and outside (magenta) the tube. The maximum amplitude of $v_z$ inside the flux tube (normalised by the largest value) and at the boundary, are shown in the middle panels (b) and (d). The blue dots indicate the maximum amplitude of $v_z$ inside the flux tube for various modes while the green dots shows the maximum amplitude at the tube boundary. Panels (c) and (f) show the absolute value of the ratio of the maximum amplitude of $v_z$ at the boundary of the waveguide and the maximum value attained inside the waveguide. On the horizontal axis of the panels (b), (c), (e) and (f) we denote the modes that correspond to different value of the radial wavenumber. }
\label{fig:f2}
\end{figure*}

\begin{figure*}
\gridline{\fig{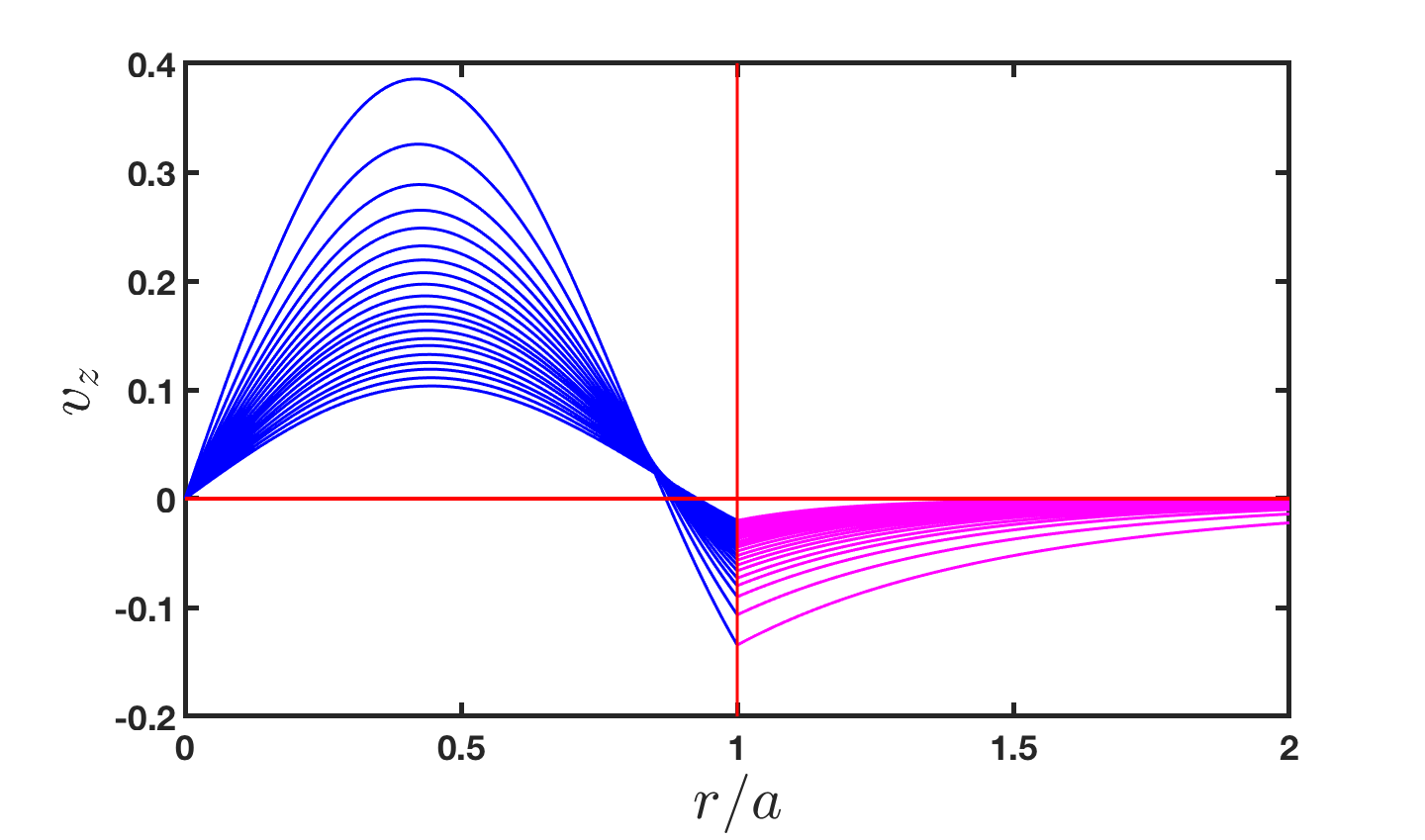}{0.35\textwidth}{(a)}
          \fig{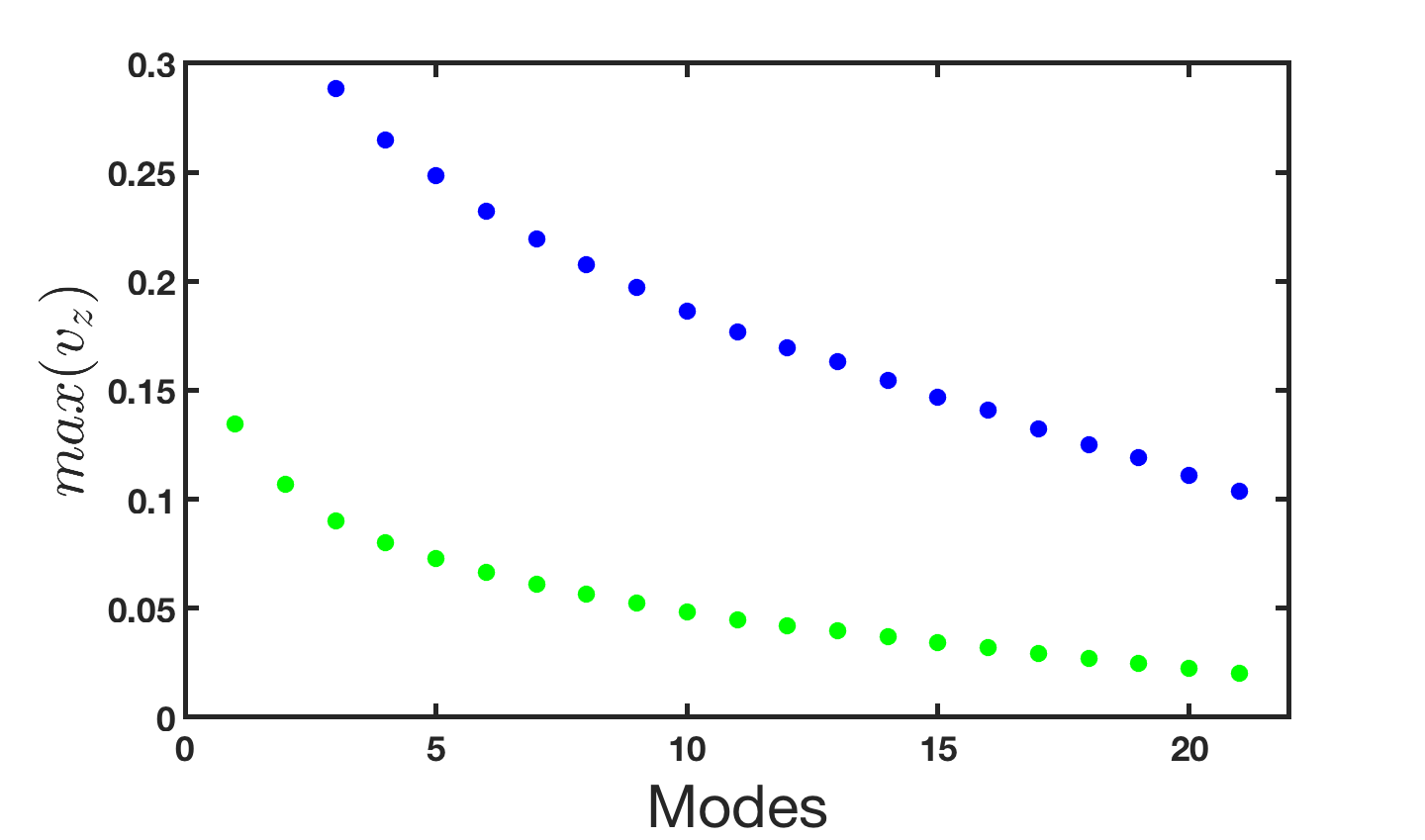}{0.35\textwidth}{(b)}
          \fig{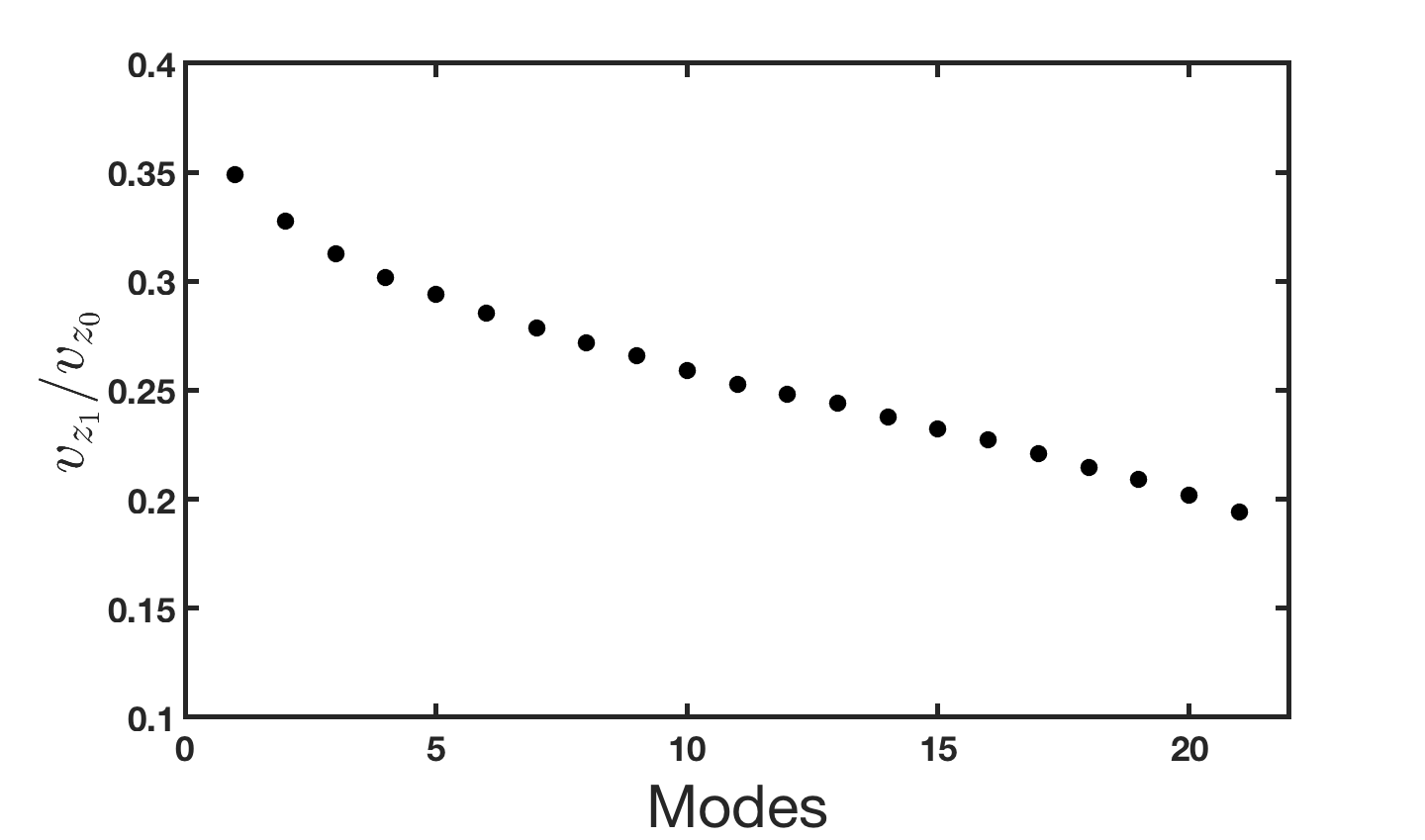}{0.35\textwidth}{(c)}
          }
\gridline{\fig{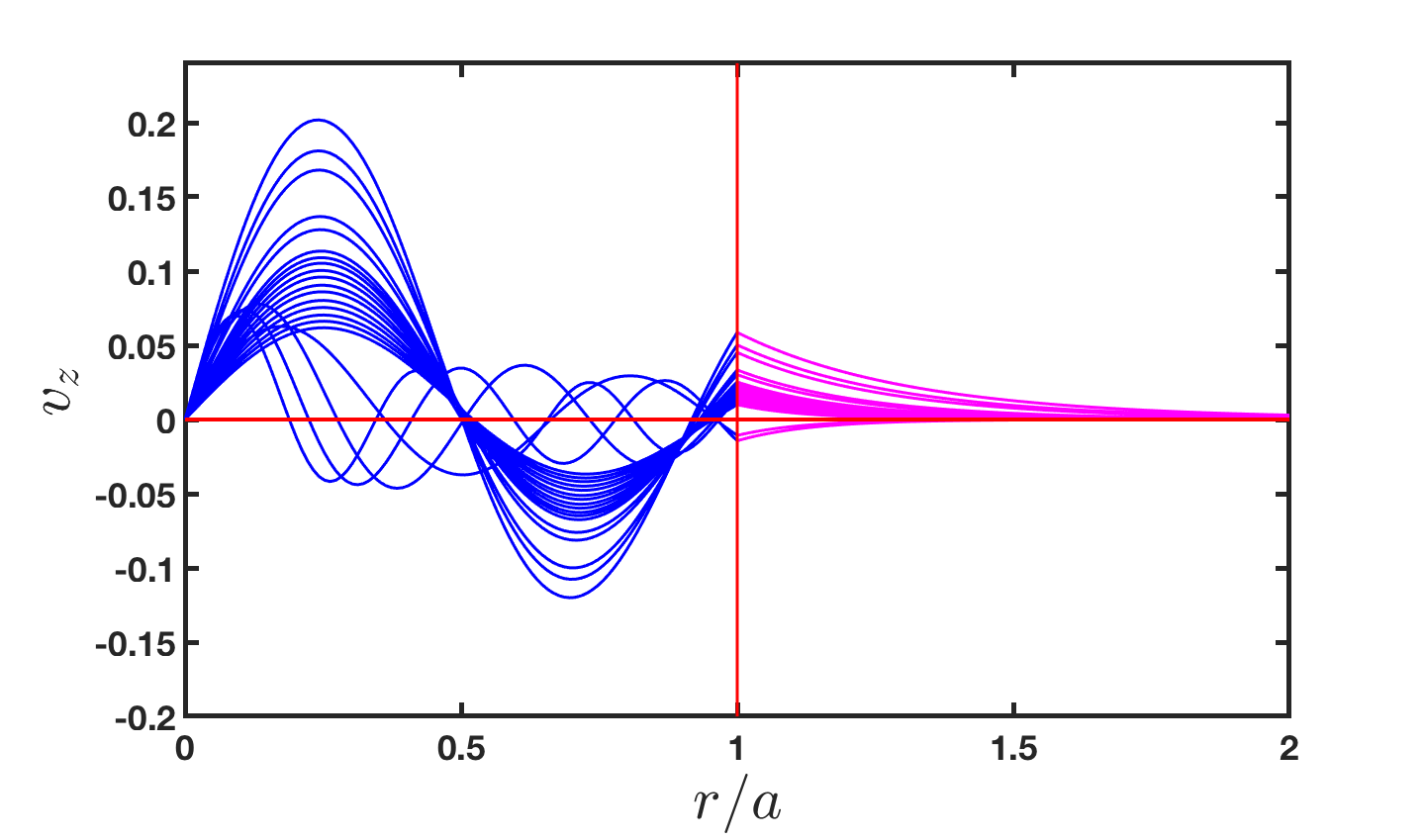}{0.35\textwidth}{(d)}
          \fig{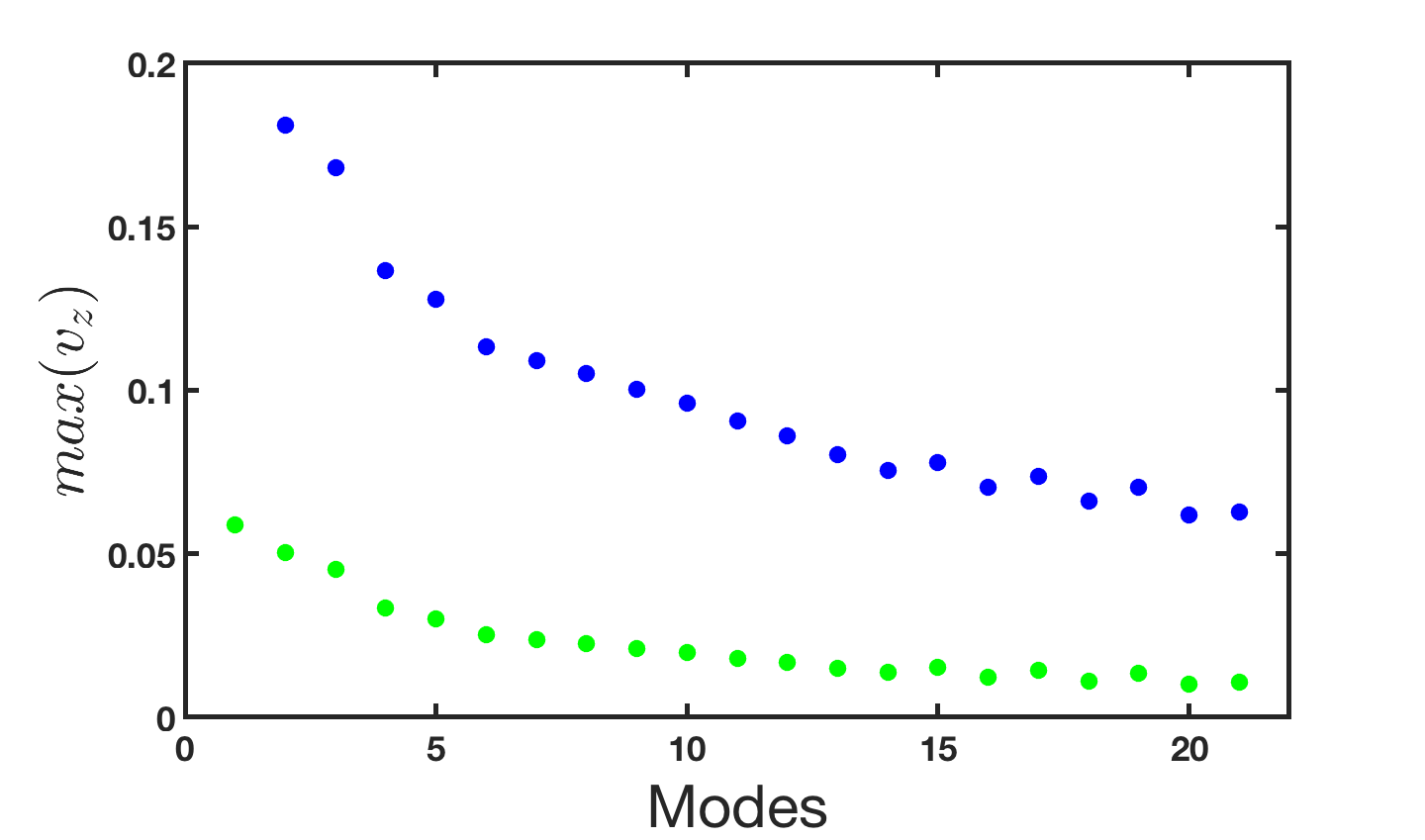}{0.35\textwidth}{(e)}
          \fig{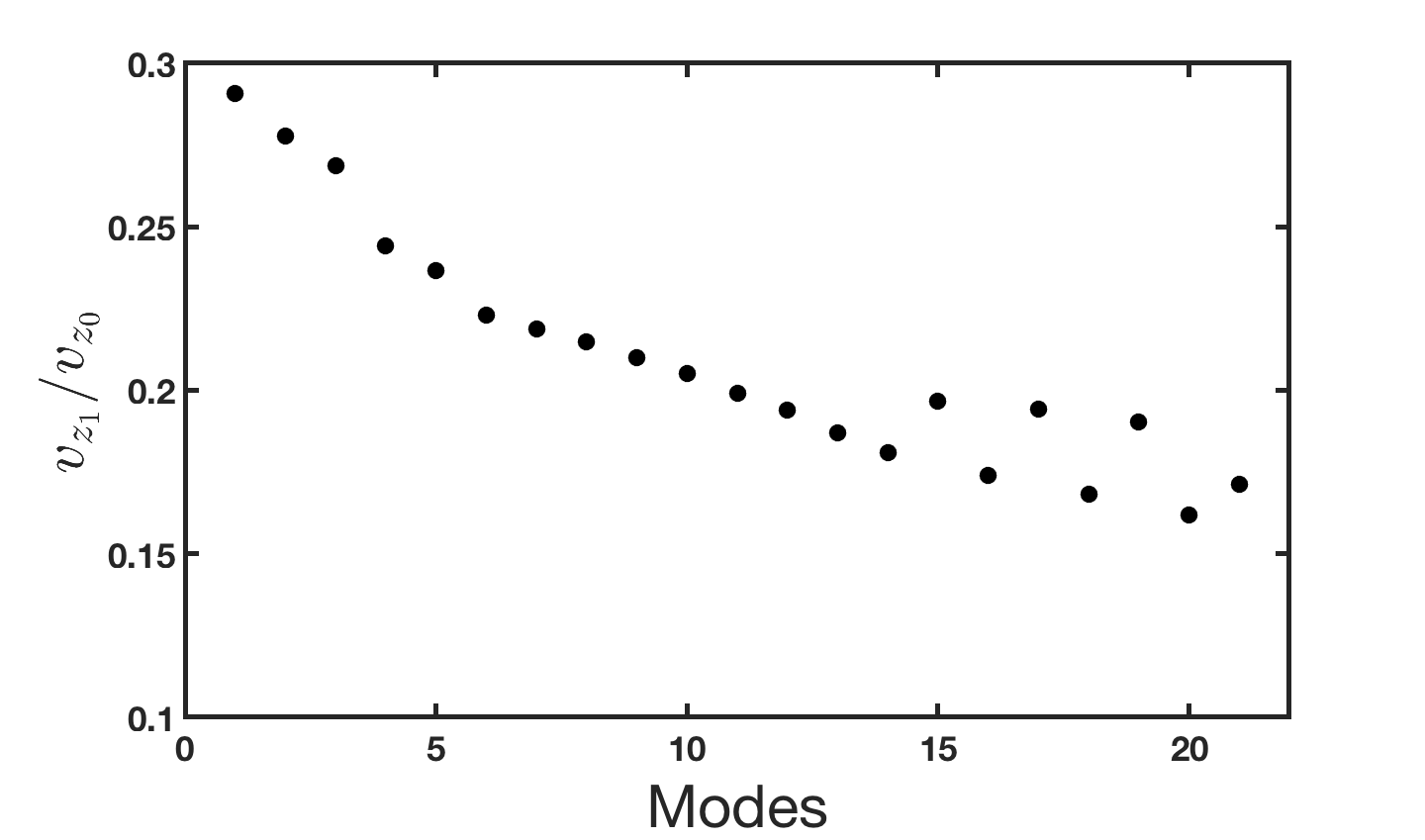}{0.35\textwidth}{(f)}
          }
\caption{The same as Figure \ref{fig:f2} but here we plot the amplitude of the $z$-component of the velocity for slow body kink modes under photospheric conditions. }
\label{fig:f3}
\end{figure*}

The radial variation of the $z$-component of the velocity for slow body sausage and kink modes are shown in Figures (\ref{fig:f2}) and (\ref{fig:f3}), panel (a) and (d), respectively. The cases corresponding to $an_i<4$ and $an_i>4$ are shown separately for simplicity. The value of 4 has been chosen to make sure that in the case of both Bessel functions $J_0$ and $J_1$ have only one zero.  

It is clear that in the case of both slow sausage and kink modes the amplitude of the $z$-component of velocity near the boundary of the waveguide ($r/a\approx 1$) under photospheric conditions is very small, and the maximum occurs somewhere inside the waveguide. This result also suggest that in the case of these waves the longitudinal perturbation at this boundary is very small. In order to evidence the relative difference between the maximum of various body modes and the amplitude of $v_z$ at the boundary we show the location of these maxima in panels (b) and (e) by blue and green dots for waves corresponding to different value of the radial wavenumber. It is clear that the fundamental sausage and kink modes have the largest amplitude inside the cylinder, while higher overtones have smaller amplitude with increasing the radial wavenumber. The same pattern is followed by the amplitude of $v_z$ at the boundary, where the amplitude of various overtones are close to zero. We note here that under photospheric conditions there are no fast body modes but there are fast surface modes and these modes are not consistent with the ones present in our model. 

\begin{figure*}
\gridline{\fig{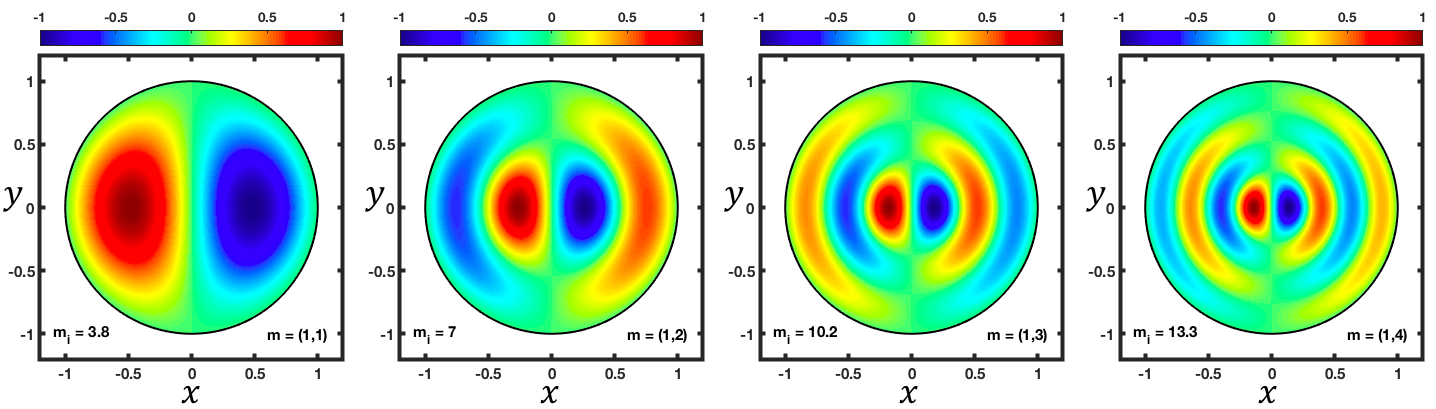}{1.0\textwidth}{}
           }
     \gridline{\fig{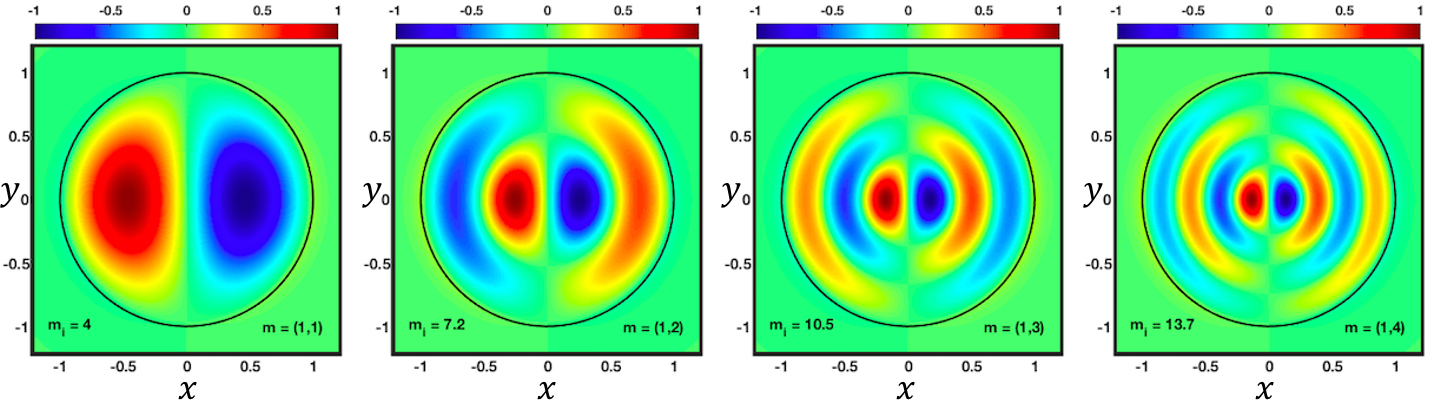}{1.0\textwidth}{}
           }
\caption{The polar representation of kink modes in a cylindrical waveguide. The  top row shows different branches of the kink modes obtained by assuming $v_z=0$ on the boundary of the cylindrical waveguide. The  bottom row shows the same branches of the kink modes obtained by solving the whole dispersion relation. The labels $(m= (x,y))$ represent the type of mode and branch, e.g. $x=1$ refers to the kink mode and $y=2$ refers to the second branch. 
\label{fig:f5}}
\end{figure*}

\begin{figure}
	\begin{center}
	\mbox{\includegraphics[scale=0.35]{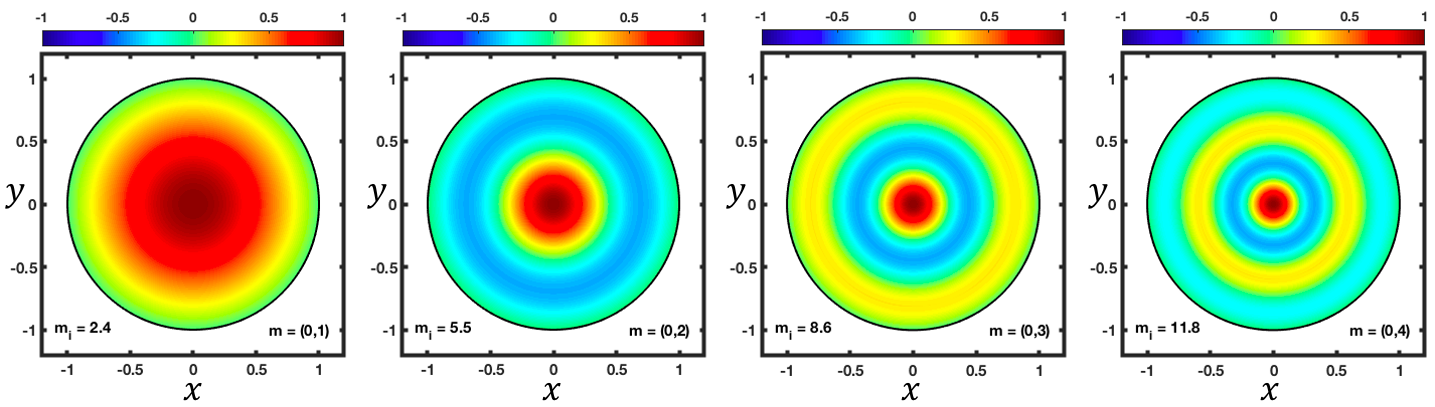}}
	\mbox{\includegraphics[scale=0.35]{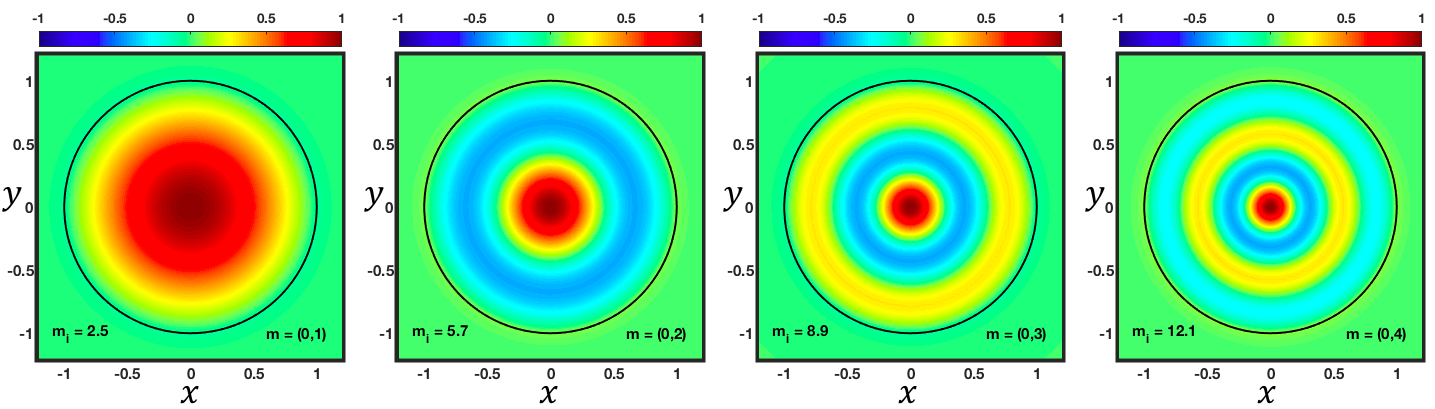}}
	\end{center}
	\caption{The same as Figure (\ref{fig:f5}) but for sausage modes.}
	\label{fig:ch_4_f4}
\end{figure}

Figures \ref{fig:f5} and \ref{fig:ch_4_f4} display the 2D polar plot of the $v_z$ component of velocity as obtained as a result of solving the dispersion relation (Eq. \ref{eq:4}) (upper rows) and the solution obtained by imposing the condition $v_{zi}=0$ at the boundary (lower panels), where the expression of $v_{zi}$ is given by Eq. (\ref{eq:8}). For illustration we show the first four branches of kink (Figure \ref{fig:f5}) and sausage modes (Figure \ref{fig:ch_4_f4}).  The modes $m$ are labelled by two numbers $(x,y)$, where the first number ($x$) denotes the azimuthal wavenumber ($x=0$ and $x=1$ correspond to the sausage and kink modes), while the second number ($y$) denotes the different branches e.g. $y=1$ corresponds to the first branch, $y=2$ to the second, etc. Figures \ref{fig:f5} and \ref{fig:ch_4_f4} show, again, that our simplified approach is consistent with the standard assumption which requires the continuity of the radial velocity component and the total pressure for all modes. The white regions in the external regions of the waveguide in the upper panels denote that no solution is sought as these regions are not needed for the analysis.

The above results can be easily understood even from mathematical point of view. Let us introduce a new variable in Equation (\ref{eq:6}) such that 
\begin{equation}
    \xi=\frac{a-r}{a},
    \label{eq:4.1}
\end{equation}
where $a$ is the radius of the tube. As a result, in the new variable, the centre of the cylinder is situated at $\xi=1$ and the tube's boundary corresponds to $\xi=0$. In the new variable, the governing equation for the total pressure reads
\begin{equation}
    \frac{d^2 P_i}{d \xi^2}+\frac{1}{a^2(1-\xi)}\frac{d P_i}{d \xi}+\left(a^2n_i^2-\frac{n^2}{(1-\xi)^2}\right)P_i=0.
    \label{eq:4.2}
\end{equation}
This equation is still a Bessel differential equation whose solution bounded at the origin can be written as
\begin{equation}
    P_i\propto (1-\xi)J_{\sqrt{n^2+1}}[an_i(\xi-1)],
    \label{eq:4.3}
\end{equation}
where $n$ takes the value of 0 and 1 for sausage and kink modes, respectively. Since we are interested in the behaviour or the total pressure near the boundary, we can expand this solution into series  about $\xi\approx 0$ and the approximate solutions for the total pressure of the sausage ($P_i^s$) and kink ($P_i^k$) modes become
\begin{equation}
    P_i^s=J_1(an_i)-J_0(an_i)an_i\xi+{\cal O}(\xi^2)
    \label{eq:4.4}
\end{equation}
and
\begin{equation}
    P_i^k=J_{\sqrt{2}}(an_i)-\left[J_{\sqrt{2}+1}(an_i)+(\sqrt{2}+1)J_{\sqrt{2}}(an_i)\right]\xi+{\cal O}(\xi^2)
    \label{eq:4.5}
\end{equation}
or using the trigonometric representation of the Bessel functions
\begin{equation}
    P_i^s\approx -\sqrt{\frac{2}{\pi an_i}}\left[\cos \left(an_i+\frac{\pi}{4}\right)-an_i\left(1-\frac{r}{a}\right)\cos\left(an_i-\frac{\pi}{4}\right)\right]
\end{equation}
and
\[
P_i^k\approx \sqrt{\frac{2}{\pi an_i}}\cos\left(an_i-\frac{\sqrt{2}\pi}{2}-\frac{\pi}{4}\right)\left[1-(\sqrt{2}+1)\left(1-\frac{r}{a}\right)\right]-
\]
\begin{equation}
\sqrt{\frac{2}{an_i}}\left(1-\frac{r}{a}\right)\cos\left[an_i-\frac{(\sqrt{2}+1)\pi}{2}-\frac{\pi}{4}\right].
\label{eq:4.7}
\end{equation}
Once the solution for the total pressure is known, the $z$-component of the velocity can be determined with the help of Equation (\ref{eq:8}). These expressions clearly show that near the boundary of the cylindrical waveguide the magnitude of the solutions are indeed very small ($\approx {\cal O}(an_i)^{-1/2}$), as shown in Figures (\ref{fig:f1}) and (\ref{fig:f2}), therefore the use of the approximation $v_z=0$ on the boundary is fully justified.

\section{Conclusions}\label{sec:5}

The determination of the nature and properties of waves propagating in solar magnetic structures is one of the key ingredients in performing plasma and field diagnostics using seismological techniques. However, the determination of these properties involve treating the magnetic waveguides as an ideal environment with high degree of symmetry that allows the determination of dispersion relations. In reality, however, these waveguides are rather irregular, and the cross-sectional shape cannot be approximated by circles and/or ellipses.

In this paper we examined the validity of a simplifying method to determine the eigenvalues of slow body waves propagating under photospheric conditions and this has been performed using a cylindrical symmetry for which the dispersion relation can be derived. Our results suggest that the eigenfrequency of slow body waves and their dispersive variation can be approximated in a  simpler, yet fairly accurate way by imposing the condition that the longitudinal component of the velocity vanishes at the boundary of the waveguide. Our approach was tested by comparing our results to the solutions of the full dispersion relation. We have clearly shown that the percentage error by using this approximation is very small (less than 1\%) and the error tends to diminish with increasing the order of the modes. 

It is worth mentioning that the solutions we obtained show a gap for long-wavelength because this approximation fails near the characteristic speeds, as at these values the argument of the Bessel function becomes zero or infinity, meaning that these speeds are degenerate values of the system. Therefore, our approximate method does not work in the thin flux tube limit. In addition our approximation cannot be applied to surface waves in sunspots because surfaces waves are evanescent within the umbra and have maximum amplitude at the umbra/penumbra boundary.  

The main advantage of determining the eigenvalues of slow body modes (sausage and kink) using this approximation is that the it removes the necessity of one of the key conditions needed to derive the dispersion relation, i.e. the continuity of the radial component of the velocity. Instead, reliable solutions can be obtained by solving a Helmholtz-like differential equation with Dirichlet boundary conditions. This approximation also allows us to more easily determine the eigenvalues of various modes in magnetic waveguides with cross-sections of arbitrary shape and further validates the sunspot umbra modelling of \cite{Marco2022} and \cite{Albidah2022} who made the vertical velocity or density/pressure perturbations zero at the umbra-penumbra boundary to be consistent with observational data (it was observed that the oscillations decayed very rapidly at the boundary region). This also had the added advantage of making the computations of the eigenmodes using the observed irregular cross-sectional shapes more straightforward, since matching the transverse velocity and pressure perturbations for a magnetic waveguide with a complex cross-sectional shape is a non-trivial problem.

\acknowledgments
AA acknowledge the Deanship of Scientific Research (DSR), King Faisal University, Al-Hassa, KSA for the financial support under Raed Track (grant No.RA00029). 
\bibliography{udatefile}{}
\bibliographystyle{aasjournal} 

\end{document}